\documentclass[pra,notitlepage]{revtex4-1}
\usepackage{hyperref}
\usepackage{amsmath}
\usepackage{amssymb}
\usepackage{graphicx}

\newcommand{\bra}[1]{\ensuremath{\langle #1|}}
\newcommand{\ket}[1]{\ensuremath{|#1\rangle}}
\newcommand{\ketbra}[2]{\ket{#1}\bra{#2}}

\newcommand{\tr}{\mathrm{Tr}}

\newcommand{\Op}[1]{\ensuremath{{\cal #1}}}
\newcommand{\Id}{\Op{I}}

\newcommand{\Wojcik}{W\'ojcik{}}
\newcommand{\Pavicic}{Pavi\v{c}i\'{c}{}}
\newcommand{\PP}{ping-pong }
\newcommand{\ie}{i.e.}

\newcommand{\doiLink}[1]{DOI: \href{http://dx.doi.org/#1}{#1}}

\begin{document}

\title{A general scheme for information interception in the ping pong protocol}
\date{07/06/2016}
\author{Piotr Zawadzki}
\affiliation{Institute of Electronics, Silesian University of Technology,
Akademicka 16, 44-100 Gliwice, Poland}
\email{Piotr.Zawadzki@polsl.pl}


\author{Jaros{\l}aw Adam Miszczak}
\affiliation{Institute of Theoretical and Applied Informatics, Polish Academy of
Sciences, Ba{\l}tycka 5, 44-100 Gliwice, Poland}
\email{miszczak@iitis.pl}

\begin{abstract}
The existence of an undetectable eavesdropping of dense coded information
has been already demonstrated by \Pavicic{} 
for the quantum direct communication based on the ping-pong paradigm.
However, a)~the explicit scheme of the circuit is only given and no design rules are provided,
b)~the existence of losses is implicitly assumed, 
c)~the attack has been formulated against qubit based protocol only and 
it is not clear whether it can be adapted to higher dimensional systems.
These deficiencies are removed in the presented contribution.

A~new generic eavesdropping scheme built on a~firm theoretical background is proposed.
In contrast to the previous approach, it does not refer to the properties of the vacuum state,
so it is fully consistent with the absence of losses assumption.
Moreover, the scheme applies to the communication paradigm based on signal particles of any dimensionality.
It is also shown that some well known attacks are special cases of the proposed scheme.
\end{abstract}

\maketitle

\section{Introduction}

Quantum direct communication (QDC) aims at provision of confidentiality without resorting to classic encryption.
This is in contrast to quantum key distribution (QKD) technique,
as no shared key is established and quantum resources take over its role.
In QDC, similarly to QKD, its is assumed that legitimate parties can communicate
over open and authenticated classic channel.

The roots of QDC can be traced out to the QKD protocol of Long and Liu~\cite{PhysRevA.65.032302}
that, after slight modification proposed as the two-step protocol~\cite{PhysRevA.68.042317}, can be considered the first
protocol of this kind.
The \PP protocol~\cite{PhysRevLett.89.187902} is another QDC scheme which is easier
to implement at the price of lesser security margin and capacity.
These initial works exploited the entanglement of EPR pairs to protect transmission of sensitive information.
Ideas of these proposals have been further adapted to higher dimensional 
systems~\cite{OptCommun.253.15,OptCommun.262.134,QuantumInfProcess.10.189,QuantumInfProcess.11.1419}
and/or modified to enhance capacity via dense coding~\cite{PhysRevA.69.054301,PhysRevA.71.044305}.
The entanglement is a very fragile quantum resource and its handling is technically challenging.
This motivated the work towards exploiting quantum uncertainty, a resource used by most QKD protocols.
The first single-photon QDC protocol proposed by Deng and Long~\cite{PhysRevA.69.052319}
has been recently demonstrated experimentally~\cite{DL04-NatureandScience}.
The LM05 protocol~\cite{PhysRevLett.94.140501} is the other worth noting proposal of this kind.
The history of the development and the review of the early QDC proposals can be found in~\cite{FrontPhysChina.2.251}.

QDC protocols offer different level of security which usually results from the trade off
between practical feasibility and type of quantum resources available to communicating parties.
QDC protocols which process particles in blocks~\cite{PhysRevA.68.042317,OptCommun.253.15}
can be parametrized in such a way that probability of revealing sensitive information is arbitrarily small.
However, they assume that legitimate parties have long term quantum memory.
Protocols that process particles individually are quasi secure~\cite{PhysLettA.372.3953,FrontPhysChina.2.251,QuantumInfProcess.12.149}.
Quasi security means, that before eavesdropping detection, which is inevitable for long sequences,
part of the sensitive information may be revealed to the eavesdropper.
QDC is a more versatile cryptographic primitive than QKD.
In fact, QDC protocols can be used as engines for key agreement.
Any key agreement protocol executed in a private channel provided by a QDC
protocol offering unconditional security has security comparable with QKD.
Also quasi secure QDC protocols can realize unconditionally secure QKD.
However, in this case QDC phase delivers shared sequence that is partially known to the eavesdropper.
By the appropriate postprocessing \ie, privacy amplification,
the eavesdropper's knowledge on the resulting sequence can be reduced to arbitrary small
value provided that his information on the initial sequence is less that mutual information of the legitimate parties.
The realization of the QKD via QDC can be potentially more efficient as the basis reconciliation step,
that severely plaques efficiency of many QKD protocols, can be avoided~~\cite{PhysRevA.84.042344,PhysRevA.88.062302,SciRep.4.4936}.
Protocols of this type are referred as deterministic QKD and some of them have been recently
experimentally demonstrated~\cite{PhysRevLett.96.200501,SciRep.6.20962}.

This paper is devoted to the analysis of the (in)security of the ping-pong
protocol~-- an entanglement based QDC scheme~\cite{PhysRevLett.89.187902}.
Quasi security is provided only for perfect quantum channels~\cite{PhysLettA.372.3953} and 
the scheme becomes insecure
when losses~\cite{PhysRevLett.90.157901} and/or communication errors 
and imperfection of devices are taken into account~\cite{ChinesePhys.16.277}.
Protocol offers capacity of single bit per protocol cycle
because the authenticity of the shared EPR pair is verified only by a measurement in a single
basis.
This limits the available encoding to phase flips.
Possible capacity enhancement via dense coding leads to undetectable information
leakage as demonstrated in~\cite{PhysRevA.68.042317} and usage of mutually unbiased
bases in control measurements is required to preserve quasi security of 
the communication~\cite{PhysRevA.69.054301}.
In our previous work we have proved that
this observation also holds for the qudit based protocol and that detection
probability depends on the number of bases used in the control mode~\cite{QuantumInfProcess.11.1419,QuantumInfProcess.12.569}.
Anyway, no explicit attack transformation has been given in the aforementioned papers.
The present contribution is motivated by the appearance of the circuit~\cite{PhysRevA.87.042326}
(further it will be referred as P-circuit)
capable to undetectable intercept information transmitted in the qubit based ping-pong protocol
with the following configuration: quantum channel is perfect, legitimate parties use single basis
for control measurements, information is dense coded.
In other words~-- the instantiation of the attack forecast in~\cite{PhysRevA.68.042317}.
Although P-circuit is applicable to perfect channels, it assumes the appearance of the vacuum states
in the eavesdropper's ancilla.
In consequence, it does not well fit to the existing analyses.
Shortly after its appearance, a control mode that address detection of this specific 
circuit has been proposed~\cite{IntJQuantumInf.S0219749915500525}.

We propose a generic scheme for construction of attacks
that permit undetectable eavesdropping 
under the same assumptions: quantum channel is perfect, control measurements are executed in a  single basis, 
sensitive information is dense coded.
Thus our contribution can be considered as the generalization of the result given
in \cite{PhysRevA.87.042326}.
The presented method is applicable to systems of any dimension so it can be used to construct
a plethora of new transforms.
Using introduced generalization we also demonstrate the equivalence of the attack from
\cite{PhysRevA.87.042326} and CNOT operation.
In consequence, we claim that there is no need for construction of specific 
control modes as in~\cite{IntJQuantumInf.S0219749915500525},
because any control mode able to detect CNOT operation is also able to detect circuit proposed
in \cite{PhysRevA.87.042326}.
We do \emph{not} propose the attack that is undetectable by control measurements in unbiased bases.
In fact, we think that the opposite is true~-- control measurements in mutually unbiased
bases are sufficient to statistically detect coherence break of the shared entangled
state and, that way, reveal the presence of the eavesdropper~\cite{QuantumInfProcess.12.569}.

The paper is organized as follows. 
In Section 2 we provide notation and concepts used in the text. 
Section 3 presents the main contribution.
In particular we provide a general bit-flip detection scheme, 
demonstrate its equivalence with the existing approaches 
and introduce an attack on the qudit-based protocol. 
In Section 4 we summarize the presented work.

\section{Preliminaries}

\subsection{Ping-Pong protocol}\label{sec:ping-pong}

Communication protocol described below is a ping-pong paradigm variant analysed in~\cite{PhysRevA.87.042326}.
Compared to the seminal version~\cite{PhysRevLett.89.187902}, it differs only on the encoding operation~--
the sender uses dense coding instead of phase flips.
The remaining elements of the communication scenario are left intact.

\begin{figure}
\centering
\includegraphics{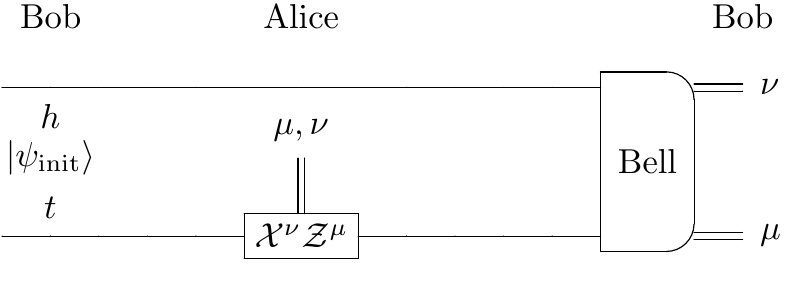}
\caption{The schematic diagram of a message mode in the \PP protocol.}
\label{pzawadzki:fig:pp-message}
\end{figure}

Bob starts the communication process by creation of EPR pair%
\footnote{The assumed initial state is the same as in~\cite{PhysRevLett.89.187902,PhysRevA.87.042326} to maintain compatibility
of mathematical expressions. For the qudit version of the protocol, considered is section~\ref{sec:generic-bitflip},
it is assumed that Bob starts from the generalization of $\ket{\Phi^+}=(\ket{0_h}\ket{0_t}+\ket{1_h}\ket{1_t})/\sqrt{2}$.}
\begin{equation}\label{pzawadzki:eq:perf-initial}
\ket{\Psi_\mathrm{init}}=\ket{\Psi^-}=\left(\ket{0_h}\ket{1_t}-\ket{1_h}\ket{0_t}\right)/\sqrt{2}
.
\end{equation}
Then he sends one of the qubits, further referred to as the~signal/travel qubit, to Alice.
Alice can in principle encode two classic bits $\mu$, $\nu$
applying unitary transformation $\Op{A}_{\mu,\nu}=\Op{X}^\mu \Op{Z}^\nu$
where 
$\Op{X}=\ketbra{1}{0}+\ketbra{0}{1}$,
$\Op{Z}=\ketbra{0}{0}-\ketbra{1}{1}$ are bit-flip and phase-flip operations, respectively.
The signal particle is sent back to Bob, 
who detects applied transformation by a collective measurement of both 
qubits~(\figurename~\ref{pzawadzki:fig:pp-message}).

\begin{figure}
\centering
\includegraphics{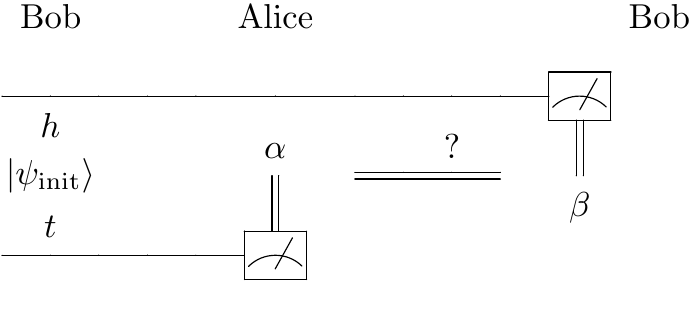}
\caption{The schematic diagram of a control mode in the \PP protocol.}
\label{pzawadzki:fig:pp-control}
\end{figure}

Passive eavesdropping is impossible. Eve has access only to the travel qubit
which before and after encoding looks like maximally mixed state.
Unfortunately, the described communication scenario is vulnerable to the intercept-resend attack
and Alice have to check whether the received qubit is genuine.
As a~countermeasure, Alice 
measures the~received qubit in computational basis ($\ket{0}$, $\ket{1}$)
in randomly selected protocol cycles
and asks Bob over authenticated classic channel 
to do the same with 
his qubit~(\figurename~\ref{pzawadzki:fig:pp-control}).
Her measurement causes the collapse of the shared state~\eqref{pzawadzki:eq:perf-initial}.
The perfect (anti)correlation of the outcomes is preserved only if the qubit measured by Alice
is the same one that was sent by Bob.
If Eve inserts fake qubit then measured qubits are no longer correlated
and some discrepancies, that are the sign of the eavesdropping, do occur.
That way Alice and Bob can convince themselves 
with confidence approaching certainty that the quantum channel
is not spoofed, provided that they have executed a~sufficient number of control cycles.

\begin{figure}
\centering
\includegraphics{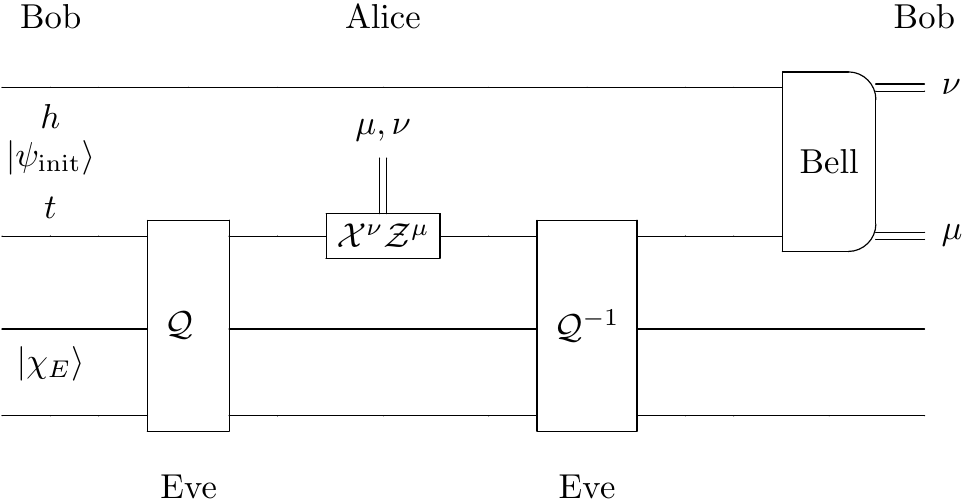}
\caption{A schematic diagram of an individual attack.}
\label{pzawadzki-1:fig:pp-attack}
\end{figure}

However, the intercept-resend attack is not the only possible way 
of active sensitive information interception.
The signal particle that travels forth and back between legitimate parties
can be the~subject of any quantum action introduced by Eve~(\figurename~\ref{pzawadzki-1:fig:pp-attack}).
Introduced coupling causes that encoding operation also modifies Eve's ancilla state
and Eve hopes to detect decipher Alice's actions by its inspection.
Actions of Eve, not necessarily unitary in the affected qubit's space,
can be described as unitary operation~$\Op{Q}$
acting in the space extended with two additional qubits,
as follows from Stinespring's dilation theorem.
The control state shared by legitimate parties then takes the form
\begin{equation}\label{eq:control-state}
\ket{\psi_{htE}}=\left(\Op{I}_h\otimes\Op{Q}\right)\left(\ket{\Psi_\mathrm{init}}\otimes\ket{\chi_E}\right)
\end{equation}
where $\ket{\chi_E}$ is some initial state of Eve's ancilla.
Eve presence is detected with probability
\begin{equation}\label{eq:control-equation}
p_\mathrm{det}(\Op{Q}) = \tr\left(
\Op{P}_{ht}\,\tr_E
\left(\ketbra{\psi_{htE}}{\psi_{htE}}\right)
\right)
\end{equation}
where projection $\Op{P}_{ht}$ depends on initial state and the considered case it is defined as
\begin{equation}
\Op{P}_{ht} = \Op{I}_{ht} - \ketbra{0_h}{0_h}\otimes\ketbra{1_t}{1_t} - \ketbra{1_h}{1_h}\otimes\ketbra{0_t}{0_t}
.
\end{equation}

\subsection{\Pavicic{} attack}\label{sec:Pavicic}

\Pavicic{}'s attack demonstrates the violation of ping-pong protocol security 
when dense coding is used.
The attack does not introduce errors nor losses in control and message mode
and it permits eavesdropping information encoded as bit flip operation.

\begin{figure}
\centering
\includegraphics{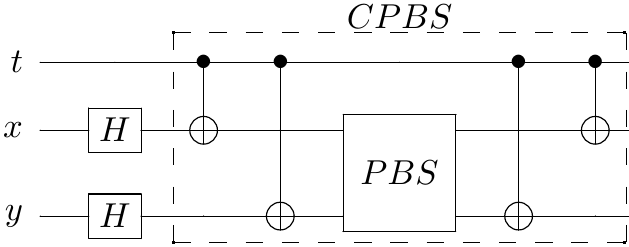}
\caption{P-circuit~$\Op{Q}_{txy}$~\cite[eq.~(2)]{PhysRevA.87.042326}.}
\label{pzawadzki:fig:cpbs}
\end{figure}

The P-circuit presented by \Pavicic{}~(\figurename~\ref{pzawadzki:fig:cpbs}) is a~result of 
a~cut and try procedure~\cite[section IV]{PhysRevA.87.042326}
applied to the \Wojcik's circuit~\cite{PhysRevLett.90.157901}.
It is is composed of two Hadamard gates followed by 
the~controlled polarization beam splitter~($CPBS$),
which is a generalization of the polarization beam splitter ($PBS$) concept.
The $PBS$ is a two port gate that
swaps horizontally polarized photons $\ket{0_x}$ ($\ket{0_y}$) entering its input to the other port $\ket{0_y}$ ($\ket{0_x}$)
on output while vertically polarized ones $\ket{1_{x}}$ ($\ket{1_{y}}$) retain in their port $\ket{1_{x}}$ ($\ket{1_{y}}$)
\ie:
\begin{subequations}
\begin{align}
PBS \ket{v_x}\ket{0_y} &= \ket{0_x}\ket{v_y} ,& \quad PBS \ket{v_x}\ket{1_y} &= \ket{v_x}\ket{1_y} ,\\ 
PBS \ket{0_x}\ket{v_y} &= \ket{v_x}\ket{0_y} ,& \quad PBS \ket{1_x}\ket{v_y} &= \ket{1_x}\ket{v_y} ,
\end{align}
\end{subequations}
where $\ket{v}$ denotes the vacuum state.
The $CPBS$ behaves as normal $PBS$ if control qubit is set to $\ket{0_t}$.
The roles of horizontal and vertical polarization are exchanged for control qubit set to~$\ket{1_t}$:
\begin{subequations}
\begin{align}
CPBS \ket{0_t}\ket{v_x}\ket{0_y} &= \ket{0_t}\ket{0_x}\ket{v_y} ,&  CPBS \ket{1_t}\ket{v_x}\ket{0_y} &= \ket{1_t}\ket{v_x}\ket{0_y}  ,\\ 
CPBS \ket{0_t}\ket{0_x}\ket{v_y} &= \ket{0_t}\ket{v_x}\ket{0_y} ,&  CPBS \ket{1_t}\ket{0_x}\ket{v_y} &= \ket{1_t}\ket{0_x}\ket{v_y}  ,\\
CPBS \ket{0_t}\ket{v_x}\ket{1_y} &= \ket{0_t}\ket{v_x}\ket{1_y} ,&  
CPBS \ket{1_t}\ket{v_x}\ket{1_y} &= \ket{1_t}\ket{1_x}\ket{v_y}  ,\\
CPBS \ket{0_t}\ket{1_x}\ket{v_y} &= \ket{0_t}\ket{1_x}\ket{v_y} ,&  
CPBS \ket{1_t}\ket{1_x}\ket{v_y} &= \ket{1_t}\ket{v_x}\ket{1_y}  .
\end{align}
\end{subequations}
Initially Eve's ancilla is initialized to the state 
$\ket{\chi_0}=\ket{v_x}\ket{0_y}$. 
The action of the P-circuit from \figurename~\ref{pzawadzki:fig:cpbs} is then described by the following formulas
\begin{subequations}\label{pzawadzki:eq:Pavicic-analysis-initial}
\begin{align}
\Op{Q}_{txy} \ket{0_t}\ket{\chi_0} &= 
\frac{1}{\sqrt{2}}\ket{0_t}\left(\ket{0_x}\ket{v_y}+\ket{v_x}\ket{1_y}\right) 
= \ket{0_t}\ket{a_E} ,\\ 
\Op{Q}_{txy} \ket{1_t}\ket{\chi_0} &=
\frac{1}{\sqrt{2}}\ket{1_t}\left(\ket{v_x}\ket{0_y}+\ket{1_x}\ket{v_y}\right) = 
\ket{1_t}\ket{d_E} .
\end{align}
\end{subequations}
For the purpose of future analysis, let us also identify actions of the circuit
under consideration onto the state 
$\ket{\chi_1}=\ket{0_x}\ket{v_y}$:
\begin{subequations}\label{pzawadzki:eq:Pavicic-analysis-orthogonal}
\begin{align}
\Op{Q}_{txy} \ket{0_t}\ket{\chi_1} &=
\frac{1}{\sqrt{2}}\ket{0_t}\left(\ket{v_x}\ket{0_y}+\ket{1_x}\ket{v_y}\right) = 
\ket{0_t}\ket{d_E} ,\\ 
\Op{Q}_{txy} \ket{1_t}\ket{\chi_1} &=
\frac{1}{\sqrt{2}}\ket{1_t}\left(\ket{0_x}\ket{v_y}+\ket{v_x}\ket{1_y}\right) = 
\ket{1_t}\ket{a_E}.
\end{align}
\end{subequations}
The control state~\eqref{eq:control-state} after entangling with Eve's ancilla reads
\begin{equation}\label{pzawadzki:eq:perf-Alice}
\ket{\psi_{htE}}=\left(\ket{0_h}\ket{1_t}\ket{d_E}+\ket{1_h}\ket{0_t}\ket{a_E}\right)/\sqrt{2}
.
 \end{equation}
This state is further used by Alice and Bob for eavesdropping check.
It is clear from \eqref{eq:control-equation} that attack 
does not introduce errors nor losses in control mode and the expected correlation of outcomes is preserved
in the computational basis.
\begin{description}
\item[Phase flip.] The phase flip encoding applied to the coupled state leads to
\begin{equation}
\ket{\psi_\mathrm{phase}} = (\Id_h\otimes \Op{Z}_t) \ket{\psi_{htE}} =
\frac{1}{\sqrt{2}}\left(\ket{1_h}\ket{0_t}\ket{a_E}-\ket{0_h}\ket{1_t}\ket{d_E}\right)
.
\end{equation}
The signal qubit is then sent back to Bob who, after disentangling on a basis of \eqref{pzawadzki:eq:Pavicic-analysis-initial},
observes
\begin{multline}
\ket{\phi_\mathrm{phase}} = 
(\Op{Q}_{txy})^{-1}\ket{\psi_\mathrm{phase}} =
\frac{1}{\sqrt{2}}\left(\ket{1_h}\ket{0_t}-\ket{0_h}\ket{1_t}\right) \ket{v_x}\ket{0_y}  =
\left[\left(\Id_h\otimes\Op{Z}_t\right) \ket{\psi_\mathrm{init}} \right] \ket{\chi_0}
.
\end{multline}
\item[Bit flip.] The bit flip operation transforms Alice's state to
\begin{equation}
\ket{\psi_\mathrm{bit}} = (\Id_h\otimes \Op{X}_t) \ket{\psi_{htE}} =
\frac{1}{\sqrt{2}}\left(\ket{1_h}\ket{1_t}\ket{a_E}+\ket{0_h}\ket{0_t}\ket{d_E}\right)
.
\end{equation}
The system state after disentangling can be deduced 
from~\eqref{pzawadzki:eq:Pavicic-analysis-orthogonal}:
\begin{multline}
\ket{\phi_\mathrm{bit}} = 
(\Op{Q}_{txy})^{-1}\ket{\psi_\mathrm{bit}} =
\frac{1}{\sqrt{2}}\left(\ket{1_h}\ket{1_t} +\ket{0_h}\ket{0_t}\right) \ket{0_x}\ket{v_y}
=
\left[ \left(\Id_h\otimes\Op{X}_t\right)\ket{\psi_\mathrm{init}} \right]\ket{\chi_1}
.
\end{multline}
\end{description}
In both cases \ie, phase flip and bit flip encodings, 
the signalling subsystem behaves as if there was no coupling with the ancilla.
However, Alice's bit flip encoding modifies Eve's register ($\ket{\chi_0}\to\ket{\chi_1}$).
The states $\ket{\chi_0}$ and $\ket{\chi_1}$ are orthogonal and perfectly distinguishable.
In consequence Eve can eavesdrop bit flip operations without introducing errors and losses
in message mode as well.

\section{Results}

This section is devoted to the analysis of the general form of the incoherent attack
shown diagrammatically in~\figurename~\ref{pzawadzki-1:fig:pp-attack}.
Each cycle of the protocol is considered to be independent on the other ones.
Consequently, the effectiveness of the attack is expressed in a fraction of eavesdropped bits per communication cycle.
Throughout the analysis it is also assumed that legitimate parties rely on control mode used
in the seminal version of the protocol.
They locally measure possessed particles in the computational basis
and verify expected correlation via the public discussion over authenticated classic channel.

\subsection{Generic bit-flip detection scheme for qubit based protocol}\label{sec:generic-bitflip}

As the control mode explores outcomes of local
measurements in computational basis for intrusion detection,
the map $\Op{Q}$ has to be of trivial form
\begin{equation}\label{pzawadzki:eq:undetectable-individual}
\Op{Q} \ket{0_t}\ket{\chi_E} \to \ket{0_t}\ket{a_E}, \quad\quad
\Op{Q} \ket{1_t}\ket{\chi_E} \to \ket{1_t}\ket{d_E}
\end{equation}
to not induce errors and/or losses in control cycles. 
It follows, that under attack,  Alice operates on the state
\begin{equation}\label{eq:control-equation-qubits-computational}
\ket{\psi_{htE}}=\frac{1}{\sqrt{2}}\left(\ket{0_h}\ket{1_t}\ket{d_E}-\ket{1_h}\ket{0_t}\ket{a_E}\right)
.
\end{equation}
Let the entangling transformation
$\Op{Q}$ additionally satisfies
\begin{equation}\label{pzawadzki:eq:undetectable-individual-bitflip}
\Op{Q} \ket{0_t}\ket{\phi_E} \to \ket{0_t}\ket{d_E}, \quad\quad
\Op{Q} \ket{1_t}\ket{\phi_E} \to \ket{1_t}\ket{a_E}
\end{equation}
for some state $\ket{\phi_E}\neq \ket{\chi_E}$.
The process of information encoding and disentangling from the ancilla
is then described by the expressions:
\begin{subequations}
\begin{multline}
\Op{Q}^{-1}\left(\Op{I}_h\otimes\Op{I}_t\otimes\Op{I}_E\right)\ket{\psi_{htE}}
=
\Op{Q}^{-1}
\frac{1}{\sqrt{2}}\left(\ket{0_h}\ket{1_t}\ket{d_E}-\ket{1_h}\ket{0_t}\ket{a_E}\right)
= \\ 
=
\frac{1}{\sqrt{2}}\left(\ket{0_h}\ket{1_t}\ket{\chi_E}-\ket{1_h}\ket{0_t}\ket{\chi_E}\right)
= \ket{\Psi^{-}}\ket{\chi_E}\,,
\end{multline}
\begin{multline}
\Op{Q}^{-1}\left(\Op{I}_h\otimes\Op{X}_t\otimes\Op{I}_E\right)\ket{\psi_{htE}}
=
\Op{Q}^{-1}
\frac{1}{\sqrt{2}}\left(\ket{0_h}\ket{0_t}\ket{d_E}-\ket{1_h}\ket{1_t}\ket{a_E}\right)
= \\ 
=
\frac{1}{\sqrt{2}}\left(\ket{0_h}\ket{0_t}\ket{\phi_E}-\ket{1_h}\ket{1_t}\ket{\phi_E}\right)
= \ket{\Phi^{-}}\ket{\phi_E}\,,
\end{multline}
\begin{multline}
\Op{Q}^{-1}\left(\Op{I}_h\otimes\Op{Z}_t\otimes\Op{I}_E\right)\ket{\psi_{htE}}
=
\Op{Q}^{-1}
\frac{-1}{\sqrt{2}}\left(\ket{0_h}\ket{1_t}\ket{d_E}+\ket{1_h}\ket{0_t}\ket{a_E}\right)
= \\ 
=
\frac{-1}{\sqrt{2}}\left(\ket{0_h}\ket{1_t}\ket{\chi_E}+\ket{1_h}\ket{0_t}\ket{\chi_E}\right)
= -\ket{\Psi^{+}}\ket{\chi_E}\,,
\end{multline}
\begin{multline}
\Op{Q}^{-1}\left(\Op{I}_h\otimes\Op{X}_t\Op{Z}_t\otimes\Op{I}_E\right)\ket{\psi_{htE}}
=
\Op{Q}^{-1}
\frac{-1}{\sqrt{2}}\left(\ket{0_h}\ket{0_t}\ket{d_E}+\ket{1_h}\ket{1_t}\ket{a_E}\right)
= \\ 
=
\frac{-1}{\sqrt{2}}\left(\ket{0_h}\ket{0_t}\ket{\phi_E}+\ket{1_h}\ket{1_t}\ket{\phi_E}\right)
= -\ket{\Phi^{+}}\ket{\phi_E}\,.
\end{multline}
\end{subequations}
As a result, the registers used for signalling are left untouched and decoupled 
but the Eve's register is flipped from $\ket{\chi_E}$ to $\ket{\phi_E}$ when Alice
applies bit-flip operation. In consequence, Eve can successfully decode
a half of the message content provided
that the detection states $\ket{\chi_E}$, $\ket{\phi_E}$  are perfectly distinguishable.
It follows that any unitary coupling transformation $\Op{Q}$
that satisfies \eqref{pzawadzki:eq:undetectable-individual} and \eqref{pzawadzki:eq:undetectable-individual-bitflip} can be used for bit flip
detection.

\subsection{Equivalence of P and CNOT circuits}

The properties of the above generic scheme 
and the P-circuit~\cite{PhysRevA.87.042326}
perfectly coincide.
As follows from \eqref{pzawadzki:eq:Pavicic-analysis-initial} and \eqref{pzawadzki:eq:Pavicic-analysis-orthogonal},
the states $\ket{\chi_0}=\ket{v_x}\ket{0_y}$ and 
$\ket{\chi_1}=\ket{0_x}\ket{v_y}$
play the role of detection states $\ket{\chi_E}$ and $\ket{\phi_E}$, respectively.
It is also clear that transformation $\Op{Q}_{txy}$ has properties claimed in
\eqref{pzawadzki:eq:undetectable-individual} and \eqref{pzawadzki:eq:undetectable-individual-bitflip}.
Thus the P-circuit can be considered as an instance of the generic scheme described
in~section~\ref{sec:generic-bitflip}.

However, the operator $\Op{Q}$ 
satisfying~\eqref{pzawadzki:eq:undetectable-individual} 
and \eqref{pzawadzki:eq:undetectable-individual-bitflip} 
can be realized in many ways.
It seems that CNOT operation acting on a single
qubit of Eve's ancilla: $\Op{Q}=CNOT_{tx}$, $\ket{\chi_E}=\ket{0_x}$, $\ket{\phi_E}=\ket{1_x}$,
$\ket{a_E}=\ket{0_x}$, $\ket{d_E}=\ket{1_x}$
is the simplest realization of the logic behind the attack.
Such version is also practically feasible as the attacks involving probes
entangled via the CNOT operation have been already proposed in the QKD context~\cite{JModOpt.53.2251,QuantInfProc.5.11}.
As a result, both, the CNOT and P circuits are equivalent in terms
of provided information gain, detectability and practical feasibility.
Consequently, there is no need for the design of control modes that  
address P-circuit in a special manner~\cite{IntJQuantumInf.S0219749915500525}.

\subsection{An attack on qudit based protocol}

The P-circuit has no straightforward generalization to qudit based version of the protocol.
In contrast, the presented approach can be adapted with ease.
Let Bob start communication process with creation of EPR pair
\begin{equation}
\ket{\beta^{(0,0)}_{h,t}}=\frac{1}{\sqrt{D}}\sum\limits_{k=0}^{D-1} \ket{k_{h}}\ket{k_{t}}\,,
\end{equation}
where $D$ is the qudit dimension.
The travel qudit  is then sent to Alice for encoding or control measurement.
In control mode, the home and travel qubits are measured in the computational basis so the projection
$\Op{P}_{ht}$ used in control equation~\eqref{eq:control-equation} takes the form
\begin{equation}
\Op{P}_{ht}=\Op{I}_{ht}-\sum\limits_{k=0}^{D-1} \ketbra{k_h}{k_h} \otimes \ketbra{k_t}{k_t}
\end{equation}
Let, by an analogy to the qubit case,
$\ket{\alpha^{(k)}_E}$ and $\ket{a^{(k)}_E}$ be the sets of $D$ orthonormal states of the ancilla
system.
These states will be further referred to as detection and probe states, respectively.
The map used by Eve must be of the form
\begin{equation}\label{pzawadzki:eq:undetectable-individual-qudit}
\Op{Q} \ket{k_t}\ket{\alpha^{(0)}_E} \to \ket{k_t}\ket{a^{(k)}_E},\quad\quad k=0,\ldots,D-1
\end{equation}
to not introduce errors in control measurements.
Let us additionally postulate that $\Op{Q}$ satisfies
\begin{equation}\label{pzawadzki:eq:undetectable-individual-bitflip-qudit}
\Op{Q} \ket{k_t}\ket{\alpha^{(m)}_E} \to \ket{k_t}\ket{a^{(m+k\bmod D)}_E}
\end{equation}
\ie,  $\Op{Q}$ advances index $k$ positions in a set of Eve's probe states.
Similarly, $\Op{Q}^{-1}$ backwards index $k$ positions:
\begin{equation}
\Op{Q}^{-1} \ket{k_t}\ket{a^{(m)}_E} \to \ket{k_t}\ket{\alpha^{(m-k\bmod D)}_E}
.
\end{equation}
Let us recall that for qudits Alice uses
\begin{equation}
\Op{Z}= \sum\limits_{k=0}^{D-1} \omega^k \ketbra{k}{k},\quad\quad 
\Op{X} = \sum\limits_{k=0}^{D-1} \ketbra{k+1 \bmod D}{k},\quad\quad
\omega=e^{j 2 \pi / D}
\end{equation}
to encode classic $\mu$, $\nu$ ``cdits'' in the following way
\begin{equation}\label{eq:qudit-dense-encoding}
\ket{\beta^{(\mu,\nu)}_{h,t}}=\Op{X}^\mu_t \Op{Z}^\nu_t \ket{\beta^{(0,0)}_{h,t}}
=\frac{1}{\sqrt{D}}\sum\limits_{k=0}^{D-1} \omega^{k \nu} \ket{k_{h}}\ket{(k+\mu \bmod D)_t}
\end{equation}
Under attack Alice applies encoding~\eqref{eq:qudit-dense-encoding} to the state coupled according to 
the~rule~\eqref{pzawadzki:eq:undetectable-individual-qudit}
\begin{multline}
\ket{\psi_{enc}}
=
\Op{X}^\mu_t \Op{Z}^\nu_t \Op{Q} \ket{\beta^{(0,0)}_{h,t}}\ket{\alpha^{(0)}_E} 
=
\Op{X}^\mu_t \Op{Z}^\nu_t\,\frac{1}{\sqrt{D}}\sum\limits_{k=0}^{D-1} \ket{k_{h}} \ket{k_t}\ket{a^{(k)}_E}
=
\frac{1}{\sqrt{D}}\sum\limits_{k=0}^{D-1}
\omega^{k \nu} \ket{k_{h}}\ket{(k+\mu \bmod D)_t}\ket{a^{(k)}_E}
\end{multline}
The travel qubit is affected by $\Op{Q}^{-1}$ in its way back to Bob
\begin{multline}
\ket{\phi_{dec}}=\Op{Q}^{-1} \ket{\psi_{enc}}
=
\frac{1}{\sqrt{D}}\sum\limits_{k=0}^{D-1}
\omega^{k \nu} \ket{k_{h}}\left(\Op{Q}^{-1}\ket{(k+\mu \bmod D)_t}\ket{a^{(k)}_E}\right)
= \\
=
\left\{\frac{1}{\sqrt{D}}\sum\limits_{k=0}^{D-1}
\omega^{k \nu} \ket{k_{h}}\ket{(k+\mu \bmod D)_t}\right\} \ket{\alpha^{(-\mu\bmod D)}_E}
\end{multline}
The expression in curly braces is an exactly the state that Bob  expects to receive when 
there is no Eve (see \eqref{eq:qudit-dense-encoding}),
so eavesdropping also does not affect the message.
At the same time, the initial state of the ancilla is moved $\mu$ positions within the set of detection states.
As a result, Eve can unambiguously identify the value of cdit $\mu$ as long as the~detection states are mutually orthogonal.

The $\Op{C}_{X}$ (Controlled $\Op{X}$) gate seems to be the simplest instance of the attack paradigm.
Let the detection and probe sets of states be the elements of the computational basis
($\ket{\alpha^{(m)}_E}=\ket{m_E}$, $\ket{a^{(m)}_E}=\ket{m_E}$)
and the ancilla is composed of the single qudit register.
The attack operation $\Op{Q}$ can be then implemented as
\begin{equation}
\Op{Q}\ket{k_t}\ket{\alpha^{(m)}_E}=  \Op{C}_{X} \ket{k_t} \ket{m_E}
= \ket{k_t} \Op{X}_{E}^{k}\ket{m_E} = \ket{k_t}\ket{(m+k\bmod D)_E} 
\end{equation}
In obvious way the requirements \eqref{pzawadzki:eq:undetectable-individual-bitflip-qudit}
regarding properties of the $\Op{Q}$ are then fulfilled.

The existence of attacks able to undetectably eavesdrop half of the dense coded information
has been already forecast in relation to qubit~\cite{PhysRevA.68.042317}, qutrit~\cite{QuantumInfProcess.10.189} 
and qudit~\cite{QuantumInfProcess.12.569} based protocol.
However, no explicit form of the attack transformation has been given.
Presented result fills this gap in and provides some general guidelines how to construct
a coupling transformation with desired properties.

\subsection{Control mode able to detect bit-flip eavesdropping}

The insecurity of the considered protocol results from inability to detect coupling $\Op{Q}_{ht}$
with the control measurements in a single basis.
Let us consider a~qubit based protocol from section~\ref{sec:ping-pong}
with control mode enhanced to measurements in two bases~-- namely
computational basis and its dual basis \ie, eigenvectors of $\Op{X}$ gate.
In the new control mode, Alice randomly selects measurement basis, performs measurement
and asks Bob to make local measurement in the same basis.
The control state~\eqref{pzawadzki:eq:perf-Alice} in the absence of coupling takes the form
\begin{equation}
\ket{\psi_{htE}}
=\frac{1}{\sqrt{2}}\left(\ket{0_h}\ket{1_t}+\ket{1_h}\ket{0_t}\right)\ket{\chi_0}
=\frac{1}{\sqrt{2}}\left(\ket{+_h}\ket{+_t}-\ket{-_h}\ket{-_t}\right)\ket{\chi_0}
\end{equation}
where $\ket{\pm}=(\ket{0}\pm\ket{1})/\sqrt{2}$ are eigenvectors of $\Op{X}$.
It follows that legitimate parties expect anticorrelation (correlation) of outcomes in the computational (dual) basis.
Under attack undetectable in the computational basis~\eqref{pzawadzki:eq:undetectable-individual},
the control equation~\eqref{eq:control-equation-qubits-computational} 
takes the following form in the dual basis
\begin{multline}\label{eq:control-equation-qubits-dual}
\ket{\psi_{htE}}
=
\frac{1}{2\sqrt{2}}\left\{
\ket{+_h} \left(\ket{d_E}-\ket{a_E}\right)
+ \ket{-_h} \left(\ket{d_E}+\ket{a_E}\right)
\right\} \ket{+_t} 
-
\frac{1}{2\sqrt{2}}\left\{
\ket{+_h}\left(\ket{d_E}+\ket{a_E}\right)
+
\ket{-_h}\left(\ket{d_E}-\ket{a_E}\right)
\right\} \ket{-_t}
.
\end{multline}
%
Alice measurement causes the collapse to one of the states in the curly braces.
It follows that Bob can obtain $\pm 1$ outcome with equal probability, what in turns renders Eve detectability.
If control bases are selected with equal probability then bit-flip attack is detected with $p_{det}=1/4$.
The above qualitative discussion addresses bit-flip attack.
The more advanced discussion on the properties of control modes based
on mutually unbiased bases and in relation to attacks of any form
can be found in~\cite{QuantumInfProcess.12.569}.
%

\section{Conclusion}

A~generic scheme that provides undetectable eavesdropping of bit-flip
operations in the seminal version of the ping-pong protocol is introduced.
It can be considered as a generalization of the P-circuit~\cite{PhysRevA.87.042326},
but in contrast, it is deduced from the very basic properties of the coupling transformation.
Moreover, the proposed scheme can be realized  without referring to the vacuum states
so it is fully consistent with the absence of losses assumption.
The CNOT gate and P-circuit are special cases of the introduced scheme
so both approaches are equivalent.
It follows, that any control mode able to detect CNOT coupling is also
able to detect the presence of the P-circuit.
The control mode based on local measurements in randomly selected unbiased bases is an example
of such procedure.
Consequently, there is no need of 
special addressing of P-circuit in 
the security analyses.
Also, the introduced scheme can be adapted to higher dimensional systems.
It can be considered as the constructive proof of the existence of attacks 
forecast in~\cite{PhysRevA.68.042317,QuantumInfProcess.10.189,QuantumInfProcess.12.569}.

\medskip

\paragraph*{Acknowledgements}
P.~Zawadzki acknowledges support from the statutory sources 
and J.~Miszczak by the Polish
National Science Center (NCN) under grant 2011/03/D/ST6/00413.

\medskip

\paragraph*{Conflict of Interest}
The authors declare that there is no conflict of interest regarding
the publication of this manuscript.


\end{document}